\documentclass{PoS}

\newcommand{\be}{\begin{eqnarray}}

\newcommand{\ee}{\end{eqnarray}}
\newcommand{\bdm}{\begin{displaymath}}
\newcommand{\edm}{\end{displaymath}}
   
\newcommand{\ba}{\begin{array}}
\newcommand{\ea}{\end{array}}

\title{Brane Inflation and Fine Tuning}

\ShortTitle{Brane Inflation and Fine Tuning}

\author{\speaker{Enrico Pajer} \\ 
        Ludwig-Maximilians-University \\ %
Department of Physics \\ %
Theresienstr. 37, D-80333 M\"unchen, Germany \\
        E-mail: \email{enrico@theorie.physik.uni-muenchen.de}}

\abstract{We argue about the importance of embedding the successful mechanism of Inflation in the context of a fundamental theory. We review some of the attempts in this direction made in the framework of String Theory. In particular we report on recent developments in Brane Inflation with emphasis on the fine tuning issue.}

\FullConference{Carg$\grave{e}$se Summer School: Cosmology and Particle Physics Beyond the Standard Models\\
		 June 30-August 11 2007\\
		 Cargese, France}

\begin{document}

%%%%%%%%%%%%%%%%%%%%%%%%%%%%%%%%%%%%%%%%%%%%%%%%%%%%%%%%%%%%%%%%%%%%%%%%%%%%%%%%%%%

\section{Motivations: a hierarchy problem}

In the last decade, observational cosmology has provided us with a large amount of new data about the universe we inhabit. Inflation is a very successful mechanism to reproduce some of these data, e.g. CMB perturbations (see Burgess lectures at this school \cite{Burgess:2007pz}). The idea is to assume a phase of accelerated expansion, in the very early history of our universe, driven by the vacuum energy of a scalar field, the inflaton, rolling down a potential. Fundamental questions arise naturally: 
\begin{enumerate}
\item Which field plays the role of the inflaton? \label{1}
\item Where does the inflaton potential come from? \label{2}
\item How does the inflaton couple to the standard model fields? \label{3}
\end{enumerate}
It is clear that these questions can be addressed only in the context of a fundamental theory. In this perspective inflation can be a bridge from very high energy (everything in between collider energies and $M_{pl}$) physics to observations. It is fair to say that the best candidate we have for a fundamental theory of nature is String Theory (ST).

Before describing specific examples of inflation in ST, we comment on a generic difficulty that one encounters when trying to embed inflation in a fundamental theory. Large field models are those for which, during the (at least) 60 e-foldings of inflation, the inflaton varies by more than one in Planckian units: $\Delta \phi \gtrsim M_{pl}$. A prominent example of this class is chaotic inflation \cite{Linde:1983gd} $V(\phi)=\frac12 m^2\phi^2$. This "simple" potential gives a good fit to the data \cite{Spergel:2006hy} provided that the inflaton mass is of order $10^{-6} M_{pl}$.

Problems arise because what seems a simple potential is indeed quite unnatural. The first issue shows up at classical level: near a minimum every potential looks like $m^2 \phi^2$ (for some $m$) as a result of an expansion in $\phi/M_{pl}$, but once we move away from it (e.g. $\Delta \phi\sim\mathcal{O}(100)$, in Planck units, to fit the data) one would expect higher terms in the $\phi/M_{pl}$ expansion to become relevant. 

Let us now suppose we have overcome this tree level fine tuning problem. At quantum level new obstacles arise: radiative corrections might drastically change the shape of the tree level potential. We know that the mass of a scalar field, contrary to a spinor mass, receives power law radiative corrections. These are quadratically divergent in the cutoff (contrary to the log divergence of spinor masses). This is an analog of the weak scale hierarchy problem involving the Higgs mass (and similar to the cosmological constant problem). 

The classical difficulty implies the need for some fine tuning of the parameters of the model (wether or not this might be justified in terms of some "selection" principle). The quantum instability or UV sensitivity seems instead to ask for a mechanism to tame the radiative corrections (as supersymmetry does for the weak scale). Such an issue can only be addressed in the context of a fundamental, UV finite theory. Finally an even stronger requirement would be that inflation becomes a generic prediction of the theory and not just one among many alternative evolutions.

This goal is still far from being reached. Several models have been proposed and many of them have, allowing for fine tuning, potentials of the slow roll type. No existing model has successfully addressed the analog of the hierarchy problem. Therefore understanding the perturbative corrections becomes essential to be able to check if a classically successful model remains so at quantum level. 

Thanks to the progress in moduli stabilization \cite{Giddings:2001yu}, type IIB has become an attractive framework for inflation model building. Some steps towards the understanding of string loop corrections in this context have been made. An example which will be relevant for brane inflation are threshold corrections to the non perturbative superpotential \cite{Baumann:2006th}. Other corrections involve the K\"ahler potential; these have been neglected in most inflationary model, basically because of the poor control we have over them. A few known explicit results come from calculations using toroidal orbifolds \cite{Berg:2005ja}. Recently a conjecture for the generalization to a Calabi Yau has been proposed \cite{Berg:2007wt}. Confirmations come from low energy effective action calculations \cite{von Gersdorff:2005bf}. Using such a proposal, it would be interesting to investigate the role of such corrections in various inflationary models.

%%%%%%%%%%%%%%%%%%%%%%%%%%%%%%%%%%%%%%%%%%%%%%%%%%%%%%%%%%%%%%%%%%%%%%%%%%%%%%%%%%%%%%%%%%%%%%%%%%%%%%%%%%%%%%%%%%%%%%%%%%%%%%%%%%%%%%%%%%%%%%%

\section {Brane Inflation}

In D brane models \cite{Dvali:1998pa}, the inflaton is given by the position of a D brane (an open string mode). The potential is then given by the Coulomb interaction with an anti D brane. The most studied model is with D3 branes in type IIB string theory. In \cite{Kachru:2003sx} it was shown that enough e-foldings are obtained if the D3- anti D3 are located in a warped throat. The anti D3 reaches fast its minimum at the tip of the throat and the D3 slowly rolls down and eventually the two annihilate. This provides an efficient reheating mechanism, by mean of which the tension of the branes is converted, after a long cascade of decays, into standard model fields.
\begin{figure}
\centering
\includegraphics[width=0.7\textwidth, height=0.3\textheight]{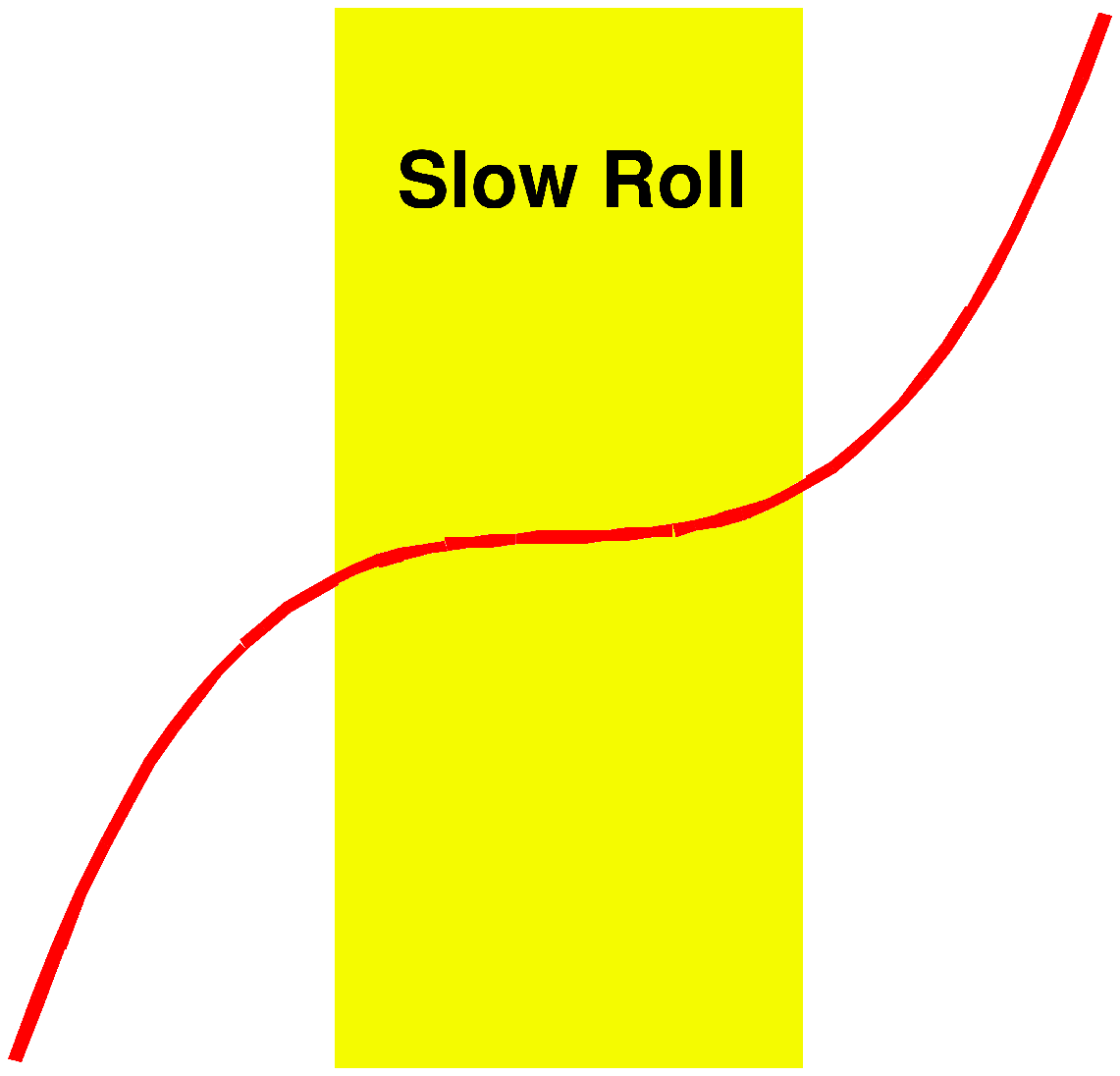}
\label{fig}
\end{figure}

In fact the brane, as any massive object, backreacts on the metric. In particular the inflaton couples (also) to the overall volume. When the volume and the other geometric moduli are fixed, i.e. the massive oscillations around their vev are integrated out, a potential is induced for the inflaton in the effective Wilsonian action. The analogy to the hierarchy problem previously discussed is clear. The Coulomb potential receives large corrections \cite{Kachru:2003sx} and becomes subleading. The inflaton acquires a mass of order the Hubble constant which prevents slow roll inflation.

It is a natural question if there are some other effects that can compete to produce a flat potential and how much fine tuning would be required. In fact, the stabilization of the overall volume requires the presence of a D7 branes (or D3 instantons) which indeed interacts with the D3 brane. This effect was calculated in \cite{Baumann:2006th} and depends on how the D7 brane is embedded in the throat. It has been found that for some embeddings \cite{Burgess:2006cb} the inflaton potential is unchanged while others \cite{Baumann:2007np} give actually rise to new terms. Unfortunately none of these terms give a contribution to the mass term, i.e. a term proportional to $\phi^2$, so that it is not possible, even with fine tuning, to cancel the large inflaton mass induced by the mixing with the stabilized volume. The traditional Coulomb driven brane inflation is therefore still lacking an explicit embedding in string theory.

On the other hand, for some value of the parameters (related to the uplifting) the inflaton potential shows an inflection point  \cite{Baumann:2007np} as in the figure. Close to this point, first and second derivatives vanish (but the potential does not) and therefore the slow roll parameters are small. The phenomenology is extremely sensitive to the shape of the potential so that it is hard to make general predictions. There are usually enough e-foldings \cite{Baumann:2007np}, especially once the multifield dynamics is taken into account \cite{Panda:2007ie}. However, in the part of the parameter space analyzed until now there seem to be no cases with both a red tilted spectrum and the right amplitude of scalar perturbations \cite{Panda:2007ie}. Comments on a possible overshooting issue were given in \cite{Itzhaki:2007nk}.

%%%%%%%%%%%%%%%%%%%%%%%%%%%%%%%%%%%%%%%%%%%%%%%%%%%%%%%%%%%%%%%%%%%%%%%%%%%%%%%%%%%%%%%%%%%%%%%%%%%%%%%%%%%%%%%%%%%%%%%%%%%%%%%%%%%%%%%%%%%%%%%

\section{Fine tuning and stringy parameters}

Typically, the fine tuning in a certain model is expressed in terms of the precise values that some effective parameters of that model have to take. However, if we want to know if such a fine tuning is possible at all, or how probable it is, we need to work in terms of fundamental, in this case stringy, parameters. The fine tuning becomes \textit{explicit} once we are able to give the values of the relevant fundamental parameters, e.g. flux numbers, such that the effective parameters, e.g. moduli vevs, take the value required by the model. 

As the number of fundamental parameters can be very large, it is easier to start scanning just a reduced set of them and see if that is enough to prove for example that a given fine tuning can be achieved. If this is not the case we should enlarge the parmeter space that we scan and try again. Eventually, either we find a solution or the search becomes technically too complicated and we cannot find an explicit fine tuning.

In the following we specialize to the fine tuning of the uplifting effective parameter in the case of brane inflation. We keep fixed all the other (effective) parameters, e.g. $W_0$ or $g_s$, and vary only the RR and NSNS fluxes $M$ and $K$ on the three cycles in the conifold. Then we show that, generically, explicit fine tuning with more than $4\%$ precision is hardly achieved. This means that to get an explicit solution we need to scan a larger parameter space.

As we said, for some values of the uplifting, the potential shows an inflection point as in the figure. For concreteness, let us suppose that the uplifting is obtained with an anti D3 brane at the tip of the throat; i.e. we have a term in the Lagragian
\be 
\frac{2T_{D3}h_0^{-1}}{V^{4/3}}\simeq\frac{2T_3e^{-8\pi K/3g_sM}} {V^{4/3}}\equiv\frac{D}{V^{4/3}}\,,
\ee
where $V$ is the volume of the 6d compact space, $g_s$ the string coupling, $T_3$ the tension of a (anti) D3 brane and $h_0$ the warp factor at the tip of the throat. Because of the quantization condition, $K$ and $M$ are integers.

An inflection point is present only for fine tuned values of $D$, let us call such a value $D_0$ \cite{Baumann:2007np}. We define $\delta$ to be the maximum percent error that we can tolerate in $D$ without spoiling inflation; in other words we want to find a solution (a vacuum of ST) such that the value for D lies inside the interval $D_0(1\pm\delta)$. If the spacing between two consecutive vacua (differing by one unit of flux) is less or equal to $D_0\delta$ we are sure that there is at least one solution that gives us inflation. In formulae
\be 
\left. 
\begin{array}{c}
T_3 \exp\left(-\frac{8\pi (K+1)}{3g_sM}\right)\\
\mathrm{or}\\
T_3 \exp\left(-\frac{8\pi K}{3g_s(M+1)}\right)
\end{array}
\right\rbrace  &
\simeq & D_0(1+\delta)\,.
\ee
It comes out that the second is the most economical possibility, i.e. it puts a milder constraint on the flux numbers; simple algebra leads to
\be 
&\frame{$\,\, M  \gtrsim  \frac{1}{3g_s\delta}\,\ln\left(\frac{T_3}{D_0}\right)$} & \label{imp}\\
& K\simeq\frac{3g_sM}{8\pi}\,\mathrm{ln}\frac{T_3}{D_0}\,.&
\ee
The important feature is that the minimum flux required is proportional to the inverse of the precision. For a numerical estimate we follow \cite{Kachru:2003sx} and take $g_s=0.1$ (a smaller value requires even larger flux numbers), $T_3=10^{-3}$ and $D_0=10^{-8}$ (the result anyway depends on the last two only logarithmically). Now the tadpole cancellation condition is
\be \label{tad}
KM=\frac{\chi}{24}\leq 75852\,,
\ee
where here the limit on $\chi$ comes from the largest known Euler characteristic for a Calabi Yau four-fold \cite{Klemm:1996ts}. Substituting \ref{imp} into \ref{tad} one obtains that $\delta\gtrsim 0.04$. This is, for example, not enough to obtain both $60$ e-foldings and a red tilted spectral index of scalar perturbation. The precision required for that is $0.1\%$ around $D_0=1.218\times 10^{-8}$ \cite{Panda:2007ie}.

Fine tuning the vacuum energy to the measured value of the cosmological constant today would be even much harder. The uplifting has then to cancel the potential with at least $\mathcal{O}(40)$ digits of precision requiring therefore a fantastically huge amount of fluxes. In the perspective of \cite{Bousso:2000xa}, this difficulty arises because we are scanning only a two dimensional space of parameters, $\{K,M\}$. Now, the way to procede is to enlarge the number of fundamental parameter to scan, for example to let other flux values vary at the same time, which changes e.g. $W_0$ and $g_s$.

%%%%%%%%%%%%%%%%%%%%%%%%%%%%%%%%%%%%%%%%%%%%%%%%%%%%%%%%%%%%%%%%%%%%%%%%%%%%%%%%%%%%%%%%%%%%%%%%%%%%%%%%%%%%%%%%%%%%%%%%%%%%%%%%%%%%%%%%%%%%%%%

\end{document}